\documentstyle[preprint,aps]{revtex}
\widetext
\draft
\tighten

\begin{document}

\title{Theorem on the proportionality of inertial and gravitational
masses in classical mechanics}

\bigskip

\author{\bf Andrew E. Chubykalo and Stoyan J. Vlaev}

\address {Escuela de F\'{\i}sica, Universidad Aut\'onoma de Zacatecas \\
Apartado Postal C-580\, Zacatecas 98068, ZAC., M\'exico}

\date{\today}

\maketitle


\baselineskip 7mm

\begin{abstract}
We considered the problem of the proportionality of inertial and
gravitational masses in classical mechanics. We found that the kinetic
energy of a material mass point $m$ in a circular motion with a constant
angular velocity around another material point $M$ depends only on its
gravitational mass.  This fact, together with the known result that the
straight line is a circumference with an infinite radius, allowed us to
prove the proportionality between the inertial and gravitational masses.
\end{abstract}


\section{Introduction} The equivalence principle
(proportionality of the inertial $m_i$ and gravitational $m_{\sl g}$
masses of a body $m$) is {\it a postulate} in classical mechanics.  It was
proven experimentally many times with a high precision. As a matter of
fact, measurements verified that $m_i$ and $m_{\sl g}$ may differ only by
a dimensionless factor constant $\eta$ \begin{equation} m_i=\eta m_{\sl
g} \end{equation}

Despite the indisputible experimental proof of this principle, a
theoretical possibility of difference between inertial and gravitational
masses is important for the theory (see e.g. [1-4]).  That is why a proof
of the equivalence principle, at least in the framework of
Newtonian classical mechanics, is of great interest.

In this report we present the proof.

\section{Theorem of equivalency}
\subsection{Kinetic energy of two bodies in a circular motion
with a constant angular velocity}

In this paragraph we discuss a well known example from classical
mechanics which showed us a way to prove the theorem. Let us
consider two bodies\footnote{In the present work we
shall consider the bodies as {\it material mass points}.} $m$ and $M$ with
inertial and gravitational masses $m_i,M_i$  and $m_{\sl g},M_{\sl g}$,
respectively. The conditions for their circular motion around the center
of inertia $(CI)$ with constant linear velocities $V_m$ and $V_M$ are:

\begin{equation}
\frac{m_i V_m^{2}}{R-x} = G\frac{m_{\sl g} M_{\sl g}}{R^{2}},\quad
\frac{M_i V_M^{2}}{x} = G\frac{m_{\sl g} M_{\sl g}}{R^{2}}
\end{equation}

In Eq.(2) $R$ is the distance between the bodies, $x$ is the distance
between the body $M$ and the $CI$, $(R-x)$ is the distance from the $CI$ to
the body $m$ and $G$ is the gravitational constant. Eq.(2) expresses
the equilibrium of the gravitational and inertial forces for both bodies.
The angular velocities $\omega_m$ and $\omega_M$ are obviously equal,
because the $CI$ always lies on the straight line passing through the
bodies $m$ and $M$.

\begin{equation}
\frac{V_m}{R-x} = \frac{V_M}{x}
\end{equation}
The system kinetic energy $K$ is:
\begin{equation}
K = \frac{m_i {V_m}^{2}}{2} + \frac{M_i {V_M}^{2}}{2},
\end{equation}
If we find the velocities $V_m$ and $V_M$ from Eqs.(2,3) and substitute in
Eq.(4) we obtain:

\begin{equation}
K= G\frac{m_{\sl g} M_{\sl g}}{2R}
\end{equation}

The fact that the inertial masses $m_i$ and $M_i$ do not enter in Eq.(5)
is quite surprising. The only thing we did was to calculate the kinetic
energy {\it without postulating the equivalency} of inertial and
gravitational masses.  The fact that the kinetic
energy does not depend on the inertial masses prompts us to consider that a
theoretical proof of the equivalence principle is perhaps possible
in classical mechanics [5].

We found\footnote {For more details see Appendix.} that the expression (5)
is also valid in a coordinate system fixed at the body $M$.

\subsection{The proof of the Theorem}
{\bf Theorem}: {\it The inertial mass $m_i$ of a body $m$ is proportional
to its gravitational mass} $m_{\sl g}$.

{\tt Proof}: $\Diamond$ We have to prove the relation (1) for {\it an
arbitrary} body $m$ and for {\it an arbitrary} velocity $V$. It means we
have to show that the constant $\eta$ {\it does not depend} on the masses
$m_i$, $m_{\sl g}$ and on the velocity $V$ of the body $m$.

First, we shall show that Eq.(1) is true for a {\it given} body
$m$ with a {\it given} velocity $V$.

So, let some body $m$ (with inertial mass $m_i$) move along a
straight line with constant velocity $V$. Its kinetic energy is:
\begin{equation}
K = \frac{m_i V^{2}}{2}
\end{equation}

From a kinematic point of view, movement with constant velocity along a
straight line and movement with constant (by magnitude) {\it linear}
velocity along a circumference of infinite radius are equivalent.

We can also consider the movement of a similar body (with the same inertial
mass $m_i$) as circumference movement around the other body $M$. The
coordinate system is fixed at the body $M$. Now we can set that the
kinetic energy and {\it linear} velocity of body $m$ be equal to that
of case (6). Nothing can prevent us from doing it. From the condition
of the equality of forces we have (see Appendix):
\begin{equation}
\frac{m_i V^{2}}{R} =
G\frac{m_{\sl g} M_{\sl g}}{R^{2}}
\end{equation}
where $m_{\sl g}, M_{\sl g}$ are gravitational masses of bodies $m$ and
$M$, $R$ is the distance between $m$ and $M$.  Expressing "$V^{2}$" from
(7) and substituting it into the formula of the kinetic energy of the body
$m$ we obtain (see Eq.(5)):
\begin{equation} K = G\frac{m_{\sl g} M_{\sl
g}}{2R}
\end{equation}

Let us now increase $M_{\sl g}$ and $R$ conserving at the same time
the value of $K$. In other words, we consider a series of independent
circular motions: in each of them the same body $m$ has the same velocity
$V$ (and, of course, the same kinetic energy $K$) and besides, the values
of $M_{\sl g}$ and $R$ increase appropriately.  In this case $M_{\sl
g}(R)$ and $R(M_{\sl g})$ are one-to-one functions.  It is obvious (see
(8)) that if $K$, $m_g$ and $G$ are constants, $M_g(R)$ and $R$
will be {\it linear dependent (proportional)} functions, i.e.  $$ M_{\sl
g}(R) = C\cdot R, $$ where $C$ is some suitable dimensional  constant (for
the considered $m_i$, $m_{\sl g}$ and $V$, of course).  After substituting
the last formula in (8) and tending $R$ to infinity\footnote{Note that if
$R$ {\it tends to infinity}, both forces (gravitational  $F_{\sl g} =
G m_{\sl g} M_{\sl g}/R^{2}$ and inertial $F_i = m_i V^{2}/R)$ tend
to zero:  $$ F_{\sl g} = G\frac{m_{\sl g}}{R} \left(\frac{M_{\sl
g}}{R}\right) = G\frac{m_{\sl g} C}{R}\rightarrow 0, \quad F_i =
\frac{m_i V^2}{R} \rightarrow 0, $$ where $m_i$, $m_{\sl g}$, $C$ and $V$
have finite constant values. A constant velocity $V$ when $R$ tends to
infinity means that the angular velocity $\omega = V/R$ tends to zero. So
we reach a situation without any forces when the body $m$ moves with a
constant velocity $V$.}  (conserving at the same time values of $K,m_{\sl
g}$ and $G$) we obtain from (8) \begin{equation} K=m_{\sl
g}\frac{G C}{2} \end{equation} where ``$G C$" has dimension
of ``$V^{2}$".  It means that (9) can be rewritten as \begin{equation}
K=m_{\sl g} \eta\frac{V^{2}}{2}, \end{equation} here $\eta$ is a
dimensionless non-defined constant.

Recalling the above-mentioned remark (equivalence between straight line
movement and the movement along the circumference of the infinite radius),
we conclude that the kinetic energy of the body moving along the straight
line with constant velocity {\it is proportional to the gravitational
mass}. Comparing (10) and (6) we obtain the equality
\begin{equation}
m_i=\eta m_{\sl g},
\end{equation}
where $\eta$ is a constant non-defined in the framework of the
above-mentioned considerations.

Now we have to prove the relation (1) (or (11)) for an arbitrary
body $m$ and for an arbitrary velocity $V$. It means we have to show that
the constant $\eta$ {\it does not depend} on the masses $m_i$, $m_{\sl
g}$ and on the velocity $V$ of the body $m$.

We shall now show that $\eta$ does not depend on the velocity $V$,
which the {\it same} body can have if it moves around $M$ but at a
different distance $R$. If we apply the same argument separately for the
velocities $V_1$ and $V_2$ of the same body $m$ we will find:

$$
K_1 = \frac{m_i V_1^2}{2}=G\frac{m_{\sl g} M_{\sl g}}{2
R_1}=G\frac{m_{\sl g} C_1}{2} = \frac{m_{\sl g} \eta_1 V_1^2}{2}
$$
\begin{equation}
{            }
\end{equation}
$$
K_2 =
\frac{m_i V_2^2}{2} =
G \frac{m_{\sl g} M_{\sl g}}{2 R_2} = G
\frac{m_{\sl g} C_2}{2} = \frac{m_{\sl g} \eta_2 V_2^2}{2} $$

\bigskip

In the expressions (12) $\eta_1=G C_1/V_1^2$ and
$\eta_2=G C_2/V_2^2$ (see Eqs.(9,10)). The constants $C_1$ and $C_2$ have
a meaning as in (9) but for the distances  $R_1$ and $R_2$ respectively.
Now from (12) follows:
\begin{equation}
m_i = \eta_1 m_{\sl g} \qquad
m_i = \eta_2 m_{\sl g}
\end{equation}
and consequently

\begin{equation}
\eta_1 = \eta_2 = \eta
\end{equation}
Eq.(14) means that the constant $\eta$ {\it does not depend} on the
velocity.

The last point is to prove that Eq.(1) is true for arbitrary masses
$m_i$ and $m_{\sl g}$. We know that (1) is true for a body $m$ with masses
$m_i$ and $m_{\sl g}$. Let us now consider another body $\widetilde{m}$
with masses $\widetilde{m}_i$ and $\widetilde{m}_g$ proportional with a
different constant $\widetilde{\eta}$.

\begin{equation}
\widetilde{m}_i = \widetilde{\eta} \widetilde{m}_{\sl g}
\end{equation}
Additivity of inertial masses gives:
\begin{equation}
m_i + m_i = 2m_i  \equiv{\cal M}_i
\end{equation}
In its turn, additivity of gravitational masses gives too:
\begin{equation}
m_{\sl g} + m_{\sl g} = 2m_{\sl g}  \equiv{\cal M}_{\sl g}
\end{equation}

But we know that $m_i = \eta m_{\sl g}$, so

\begin{equation}
{\cal M}_i = m_i + m_i = \eta m_{\sl g} + \eta m_{\sl g}
    = \eta (m_{\sl g} + m_{\sl g}) = \eta 2 m_{\sl g} = \eta {\cal
M}_{\sl g} \end{equation}

Eq.(18) shows that the masses ${\cal M}_i$ and ${\cal M}_{\sl g}$ are
proportional with the same constant $\eta$. Obviously in this particular
case the new masses ${\cal M}_i$ and ${\cal M}_{\sl g}$ are different from
the masses $m_i$ and $m_{\sl g}$.  It is clear that Eqs.(15-18) can be
generalized:

\begin{equation}
\widetilde{m}_i = b m_i \qquad
\widetilde{m}_{\sl g} = b m_{\sl g}
\end{equation}
where $b$ is an arbitrary real number. We also have :

\begin{equation}
\widetilde{m}_i = b m_i = b \eta m_{\sl g} = \eta b m_{\sl g} = \eta
\widetilde{m}_{\sl g}
\end{equation}

From Eqs.(15) and (20) follows:

\begin{equation}
\eta = \widetilde{\eta}
\end{equation}

The constant $\eta$ {\it does not depend} on the masses. Now the proof is
finished.$\Diamond$

\section{Conclusion}
The proof of the theorem of the equivalency is based on three facts . The
first is the well-known result of the Euclidean geometry that a
circumference with an infinite radius $R$ coincides with a straight line.
The second is that the kinetic energy of a body $m$ in a straight linear
motion with a constant velocity $V$ is equal to the kinetic energy of this
body in a circular motion  with a linear velocity $V$ around the body $M$.
The third fact is that the kinetic energy of two bodies in a circular
motion was expressed only with their gravitational masses.

In this way we proved that the inertial and gravitational masses are
proportional.

\bigskip
\medskip

{\large \bf Appendix}

We have introduced two rectangular coordinate systems ${\cal K}$ and
$\widetilde{\cal K}$. The inertial system ${\cal K}$ is connected with the
center of inertia ($CI$) of the bodies (material mass points). The
non-inertial coordinate system $\widetilde{\cal K}$ has its origin
$\widetilde{\cal O}$ at the body $M$ and consequently moves with respect to
the system ${\cal K}$ with a linear velocity ${\bf V}_M$ (its
magnitude $V_M$ is constant) around the point ${\cal O}$ $(CI)$ along a
circular trajectory. We have chosen the planes $\widetilde{\cal XOY}$
and ${\cal XOY}$ to coincide and consequently the axes $\widetilde{\cal
OZ}$ and ${\cal OZ}$ are parallel. The system $\widetilde{\cal K}$ also
rotates around axis of rotation $\widetilde{\cal OZ}$ with an
non-zero angular velocity $\mbox{\boldmath$\omega$}=(0,0,\omega)$. An
observer in $\widetilde{\cal K}$ will see that the body $m$ moves in the
plane $\widetilde{\cal XOY}$ along a circumference of radius $R$ around
the body $M$.

The equation of motion in the system $\widetilde{\cal K}$ is:
\begin{equation}
m_i {\bf a}_{\tt (rel)}={\bf F}_{\sl g}+{\bf F}_{in},
\end{equation}
where
$$
{\bf F}_{in}=2m_i({\bf V}_{\tt (rel)}
\times\mbox{\boldmath$\omega$})-m_i\dot{\bf V}_M+m_i \omega^2 {\bf R}- m_i
(\dot{\mbox{\boldmath$\omega$}}\times {\bf R})
$$
In Eq.(22) ${\bf a}_{\tt (rel)}$ and ${\bf V}_{\tt (rel)}$ are respectively
the acceleration and velocity of the {\it relative} motion of the body $m$
with inertial mass $m_i$ in the system $\widetilde{\cal K}$,
{\bf R} is the radius-vector of the material point $m$ in
$\widetilde{\cal K}$ with components $(x_{\tt (rel)}, y_{\tt (rel)}, 0)$,
${\bf F}_{\sl g}$ is the gravitational force.

For $\omega=const.$, we have:

\begin{equation}
a_{\tt (rel)} = \frac{V_{\tt (rel)^2}}{R}
\end{equation}

Let see now, for $\omega=const.$, may we have
\begin{equation}
{\bf F}_{in}={\bf 0} ,
\end{equation}
which could allow us to write the equation of
motion (7) in $\widetilde{\cal K}$? The eq.(24) gives two equalities:

\begin{equation}
\left\{
\begin{array}{c}
{
2m_i \omega V_{{\tt (rel)}y}
-m_i\dot{V}_{Mx}+m_i{\omega}^2 x_{\tt (rel)}=0
}\\
{
  }\\
{
2m_i \omega V_{{\tt (rel)}x} -m_i\dot{V}_{My}+m_i{\omega}^2 y_{\tt
(rel)}=0
}
\end{array}
\right.
\end{equation}
After elimination of $\omega^2$ we find:

\begin{equation}
2\omega(x_{\tt (rel)} V_{{\tt (rel)}x}+ y_{\tt (rel)} V_{{\tt (rel)}y})=
x_{\tt (rel)} \dot{V}_{My}-y_{\tt (rel)} \dot{V}_{Mx}
\end{equation}
The left part of (26) is zero, because
\begin{equation}
x^2_{\tt (rel)} + y^2_{\tt (rel)}=R^2 (= const.)
\end{equation}
and the time differentiation of (27) gives:
\begin{equation}
x_{\tt (rel)}\dot{x}_{\tt (rel)}+y_{\tt (rel)}\dot{y}_{\tt (rel)}=0=
x_{\tt (rel)}V_{{\tt (rel)}x}+y_{\tt (rel)}V_{{\tt (rel)}y}
\end{equation}
Consequently, for $\omega=const$, the eq.(24) fulfills if
the right side of eq.(26) is zero, which means that the next
equality should be true:
\begin{equation}
\left\{
\begin{array}{c}
{
x_{\tt (rel)}\dot{V}_{My}=y_{\tt (rel)}\dot{V}_{Mx}
}\\
{
  }\\
{
x_{\tt (rel)}/y_{\tt (rel)}=\dot{V}_{Mx}/\dot{V}_{My}
}
\end{array}
\right.
\end{equation}

To verify if (29) is true we shall make some considerations. First, let
define a vector ${\bf V}_{\tt (tr)}$:

\begin{equation}
{\bf V}_{\tt (tr)}={\bf V}_{M}+{\mbox{\boldmath$\omega$}}\times {\bf R}
\end{equation}
The vectors ${\bf V}_{\tt (tr)}$ and ${\bf V}_{M}$ are collinear because
the vectors ${\mbox{\boldmath$\omega$}}$ and ${\bf V}_{M}$ are both
perpendicular to the vector {\bf R}.

\begin{equation}
{\bf V}_{\tt (tr)}=k {\bf V}_{M} ,
\end{equation}
where $k$ is a scalar factor. From (30) and (31) we
have:

\begin{equation}
\left\{
\begin{array}{c}
{
k V_{Mx}=V_{Mx}-\omega y_{\tt (rel)}
}\\
{
  }\\
{
k V_{My}=V_{My}+\omega x_{\tt (rel)}
}
\end{array}
\right.
\end{equation}
If we eliminate $\omega$ in (32), will find:

\begin{equation}
\left\{
\begin{array}{c}
{
x_{\tt (rel)} V_{Mx}=y_{\tt (rel)} V_{My}
}\\
{
  }\\
{
x_{\tt (rel)}/y_{\tt (rel)}=-V_{My}/V_{Mx}
}
\end{array}
\right.
\end{equation}
The body $M$ moves around $CI$ with a linear velocity ${\bf V}_M$ which has
a constant magnitude $V_M$.
\begin{equation}
V_{Mx}^2 + V_{My}^2 = V_{M}^2 (=const.)
\end{equation}
After time differentiation of (34) we obtain:

\begin{equation}
\left\{
\begin{array}{c}
{
V_{Mx} {\dot V_{Mx}} + V_{My} {\dot V_{My}} = 0
}\\
{
  }\\
{
{\dot V_{Mx}}/{\dot V_{My}} = -V_{My}/V_{Mx}
}
\end{array}
\right.
\end{equation}
Then from (33) and (35) follows:

\begin{equation}
x_{\tt (rel)}/y_{\tt (rel)} =  {\dot V_{Mx}}/{\dot V_{My}}
\end{equation}
So we have found that the relation (36) is true. But eq.(36) is the second
equation in (29). So the eq.(29) and consequently eq.(24) fulfill.
Thus, we have shown that it is possible to choose $\omega = const.$,
for which the equation of motion in the system $\widetilde{\cal K}$ looks
like:
\begin{equation}
m_i {\bf a}_{\tt (rel)}={\bf F}_{\sl g}
\end{equation}
Now from (23) and (37) we find:
\begin{equation}
\frac{m_i V_{\tt (rel)}^{2}}{R} =
G\frac{m_{\sl g} M_{\sl g}}{R^{2}}
\end{equation}
Obviously eq.(38) is the same as eq.(7).  Note that the velocity
$V_{\tt (rel)}$ in Appendix coincides with the velocity $V$ in SECTION II.

\bigskip
\medskip

{\large \bf Acknowledgments}

We are grateful to Dr. Dharam V. Ahluvalia and  Prof. Valeri V.
Dvoeglazov for many stimulating discussions and critical comments.

\end{document}